\documentclass[]{pasj02} 

\jyear{2024}
\Received{}
\Accepted{}

\usepackage[switch,mathlines]{lineno} 

\usepackage{natbib} 
\bibpunct[:]{(}{)}{;}{a}{}{,}

\usepackage{xcolor}
\usepackage{mathpazo}
\usepackage[T1]{fontenc} 
\usepackage{appendix}

\begin{document} 

\title{Spectral evolution of the narrow emission line components in optical during the 2022 nova eruption of U Scorpii}

\author{
Katsuki \textsc{Muraoka},\altaffilmark{1}\altemailmark \email{mrok@kusastro.kyoto-u.ac.jp}
Naoto \textsc{Kojiguchi},\altaffilmark{1,2}
Junpei \textsc{Ito},\altaffilmark{1}
Daisaku \textsc{Nogami},\altaffilmark{1} 
Taichi \textsc{Kato},\altaffilmark{1} \\
Yusuke \textsc{Tampo},\altaffilmark{3,4}
Kenta \textsc{Taguchi},\altaffilmark{1,2}
Keisuke \textsc{Isogai},\altaffilmark{1,2,5} 
Arthur \textsc{Leduc},\altaffilmark{6} 
Hamish \textsc{Barker},\altaffilmark{7} \\
Terry \textsc{Bohlsen},\altaffilmark{8}  
Raul \textsc{Bruzzone},\altaffilmark{9}
Forrest \textsc{Sims},\altaffilmark{10}
James \textsc{Foster},\altaffilmark{11}
Mitsugu \textsc{Fujii},\altaffilmark{12} \\
Keith \textsc{Shank},\altaffilmark{13}
Pavol \textsc{A. Dubovsky},\altaffilmark{14}
Paolo \textsc{Cazzato},\altaffilmark{15}
Stéphane \textsc{Charbonnel},\altaffilmark{16} \\
Olivier \textsc{Garde},\altaffilmark{16}
Pascal \textsc{le Dû},\altaffilmark{16}
Lionel \textsc{Mulato},\altaffilmark{16} and
Thomas \textsc{Petit}\altaffilmark{16}
}
\altaffiltext{1}{Department of Astronomy, Graduate School of Science, Kyoto University, Kitashirakawa-Oiwake-cho, Sakyo-ku, Kyoto-shi, Kyoto 606-8502, Japan}
\altaffiltext{2}{Okayama Observatory, Kyoto University, 3037-5 Honjo, Kamogata-cho, Asakuchi-shi, Okayama 719-0232, Japan}
\altaffiltext{3}{South African Astronomical Observatory, PO Box 9, Observatory, 7935, Cape Town, South Africa}
\altaffiltext{4}{Department of Astronomy, University of Cape Town, Private Bag X3, Rondebosch 7701, South Africa}
\altaffiltext{5}{Department of Multi-Disciplinary Sciences, Graduate School of Arts and Sciences, The University of Tokyo, 3-8-1 Komaba, Meguro-ku, Tokyo 153-8902, Japan}
\altaffiltext{6}{Astronomical Ring for Amateur Spectroscopy (ARAS), 18 avenue Shakespeare, 06000 Nice, France}
\altaffiltext{7}{Rutherford street Observatory, Nelson, New Zealand}
\altaffiltext{8}{Southern Astro Spectroscopy Email Ring (SASER), Mirranook Observatory, Armidale, NSW 2350, Australia}
\altaffiltext{9}{Guyancourt Astronomical Observatory, 39 rue Léonard de Vinci, Guyancourt 78280, France}
\altaffiltext{10}{Desert Celestial Observatory, 5900 E Bell St., Apache Junction, Arizona, USA}
\altaffiltext{11}{American Association of Variable Star Observers (AAVSO), Pinon Pines Estates, Frazier Park, California 93225, USA}
\altaffiltext{12}{Fujii Kurosaki Observatory, 4500 Kurosaki, Tamashima, Kurashiki, Okayama 713-8126, Japan}
\altaffiltext{13}{Astronomical Ring for Amateur Spectroscopy (ARAS), Goldthwaite, Texas, USA}
\altaffiltext{14}{Vihorlat Observatory, Mierova 4, 06601 Humenne, Slovakia}
\altaffiltext{15}{Astronomical Ring for Amateur Spectroscopy (ARAS), Miror Astrophysical Observatory, Via Vespucci n. 33, 73010 Arnesano, Lecce,Italy}
\altaffiltext{16}{Southern Spectroscopic Project Observatory Team (2SPOT), 45, Chemin du Lac, 38690 Chabons, France}

\maketitle

\begin{abstract}

    There remains debate over whether the accretion disk survives or is entirely disrupted after the nova eruption.
    In our previous paper, Muraoka et al. (2024, PASJ, 76, 293) have photometrically demonstrated 
    that the surviving accretion disk was expanded close to the L1 point during the optical plateau stage
    and then drastically shrank to the tidal truncation radius after the optical plateau stage ended.
    To approach the clarification of the physical mechanism that drives these structural changes,
    we have then conducted systematic analyses of the spectral evolution of the narrow emission line components in optical 
    over 22 d following the optical peak during the 2022 nova eruption of U Scorpii (U Sco).
    Additionally, we present its optical spectrum in quiescence 794 d after the 2022 nova eruption. 
    We find that the single-peaked narrow components of H$\alpha$ and He \textsc{ii} 4686 appeared almost simultaneously between roughly days 6 and 8, 
    preceding the onset of the disk eclipses observed after day 11.
    This finding suggests that the nova wind near the binary system may be the primary origin of these narrow components 
    and even remained active several days after the nova eruption with a velocity of approximately 1000 $\mathrm{km\ s}^{-1}$,
    likely driving the expansion of the accretion disk until the end of the optical plateau stage. 
    While the contribution of the rotating accretion disk might be dominated by that of the nova wind in the H$\alpha$ line profile,
    the outward surface flow from the expanded disk might also contribute to these narrow features during the optical plateau stage,
    making the single-peaked narrow line profiles more pronounced.

\end{abstract}


\section{Introduction}\label{sec01:intro}

    Nova eruptions can occur in close binaries where a white dwarf (WD) accretes hydrogen-rich material from the secondary star \citep{Nova}. 
    As a layer of accreted hydrogen builds up, 
    its weight crushes the material at the base of the layer in the extreme gravity of the primary WD 
    and causes electron degeneracy \citep[chapter 11 of][]{hellier}.
    Under such circumstances, a temperature perturbation would cause the nuclear heating rate to grow faster than the cooling rate \citep{wolf2013}.
    Therefore, when the accreted layer reaches a critical mass, it undergoes a thermonuclear runaway (TNR) that burns the fuel at a rate much faster than accretion,
    and the outer layer expands to initiate mass loss after the gas pressure exceeds the degeneracy pressure, observed as a sudden brightening or "nova eruption".
    (e.g., \citealp{TNR}; chapter 11 of \citealp{hellier}; \citealp{chomiuk2021} for a recent review).

    According to \citet{hachisu2006_2}, a large part of the envelope is ejected as optically thick nova wind, 
    and the ejecta photosphere centered at the WD expands much beyond the binary separation. 
    Therefore, the whole binary system becomes obscured in the photosphere just after the optical peak.
    However, as the ejecta density decreases, the ejecta photosphere recedes toward the WD along with the decline in optical brightness, 
    and the secondary or accretion disk may emerge from the photosphere.
    Eventually, the mass-loss rate driven by the nova wind, which is still blowing from the WD, gradually decreases,
    and the WD surface emerges from the photosphere.
    On the surface, the now non-degenerate remaining envelope continues steady H-burning, 
    so the photospheric temperature becomes high, several $10^5\ \mathrm{K}$, enough for us to detect luminous soft X-ray photons, 
    and the nova enters the supersoft X-ray source (SSS) stage 
    (e.g., \citealp{mkato1994}; \citealp{sala2005}; \citealp{wolf2013}).

    While the evolution of nova eruptions has been extensively studied (e.g., \citealp{hachisu2023}; \citealp{hachisu2024} for recent studies), 
    there remains debate over whether the accretion disk survives or is entirely disrupted after the nova eruption.
    Here are some studies supporting that the accretion disk is disrupted.
    \citet{zamanov2006} have computed a viscous time scale as an approximate time for the accretion disk to be rebuilt when disrupted, 160--800 d for RS Oph, 
    and they have concluded that the cessation of the optical flickerings for several months after the RS Oph 2006 eruption 
    indicates the destruction of the accretion disk.
    \citet{worters2007} have reported that these flickerings were detected on days 241--254 after the RS Oph 2006 eruption,
    indicating the re-establishment of accretion following the destruction or severe disruption of the accretion disk \citep{darnley2017}.
    Such optical flickerings were also observed on day 8 after the U Sco 2010 eruption 
    and regarded as a sign of the re-establishment of the accretion disk \citep{worters2010},
    though \citet{munari2010} have denied this interpretation, demonstrating that the optical flickerings were absent between the following days 16 and 20.
    \citet{henze2018} have suggested that the disruption of the accretion disk might shorten the SSS stage during the 2016 nova eruption of M31N 2008-12a.
    In terms of simulations, \citet{drake2010} have demonstrated that 
    the accretion disk was entirely destroyed by the early blast wave just after the U Sco 2010 eruption occurred,
    though \citet{figueira2018,figueira2025} have demonstrated that whether the accretion disk survives or is disrupted mainly depends  
    on the relative masses of the accretion disk and the nova ejecta.
    
    On the other hand, the following studies support the hypothesis that the accretion disk survives the nova eruption.
    \citet{leibowitz1992} have described that the eclipses of V838 Her three weeks after the optical peak 
    might be due to the presence of the accretion disk.
    The detailed light curves of these optical eclipses are presented in \citet{tkato2023}.
    \citet{retter1997} have indicated the superhump phenomenon in V1974 Cyg, 
    suggesting the presence of the accretion disk no later than 30 months after the nova eruption.
    \citet{sokolovsky2008} have demonstrated that there is a possibility that the accretion disk survived the RS Oph 2006 eruption and collimated the outflows.
    \citet{darnley2017} have conducted an SED analysis of the 2015 nova eruption of M31N 2008-12a, 
    concluding that the accretion disk survived the 2015 eruption.
    Theoretically, \citet{mkato2022} have stated that the expanding nova envelope is in hydrostatic balance during the very early stage of the nova eruption, 
    and \citet{hachisu2022} have stated that a shock is generated outside the photosphere far beyond the binary system after the optical peak.
    Therefore, \citet{hachisu2025} have denied the result of \citet{drake2010} that the accretion disk was disrupted by the early blast wave.

    Moreover, \citet{hachisu2000} have conducted a modeling analysis of the optical light curve during the U Sco 1999 eruption.
    When the optical light curve reaches its peak at $V \sim 7.5$ mag from the pre-eruption magnitude $V \sim 18$
    within one day after the nova eruption occurs in U Sco, it enters the early decline stage. 
    Subsequently, it enters the optical plateau stage at $V \sim 14$ mag for $\sim$10 d, 
    which is defined as the stage when the optical magnitude temporarily remains nearly constant \citep[e.g.,][]{schaefer2011,muraoka2024}.
    \citet{hachisu2000} have suggested that 
    this optical plateau stage is mainly attributed to the surviving accretion disk irradiated by the emerging hotter WD,
    and that the accretion disk may be expanded to the L1 point due to the nova wind during this period.
    The same conclusion has also been derived by a similar multiwavelength light-curve analysis of KT Eri \citep{hachisu2025}. 
    
    Based on these theoretic studies,  
    we have estimated the primary eclipse width during the U Sco 2022 eruption over $\sim$50 d after the optical peak
    in our previous paper \citep{muraoka2024}.
    As a result, we have pointed out that the accretion disk was highly likely to survive the nova eruption.
    Furthermore, we have observationally demonstrated that 
    the surviving accretion disk was expanded close to the L1 point during the optical plateau stage
    and then drastically shrank to the tidal truncation radius 
    within just a few orbital periods\footnote{The orbital period is $\sim$1.23 d \citep{schaefer2011}.} after the optical plateau stage ended.

    Our purpose in this paper is to 
    verify whether mass loss continuously drove these structural changes even long after the optical peak of the U Sco 2022 eruption, 
    although the wind mass-loss rate drastically decreases with time after the optical peak in general \citep[e.g., figure 1 of][]{hachisu2022}.
    Especially, we focus on the spectral evolution of the narrow components of H$\alpha$ and He \textsc{ii} 4686
    because they are thought to arise either from the nova wind \citep{yamanaka2010}
    with a typical velocity of as fast as 1000--1500 $\mathrm{km\ s}^{-1}$ \citep[e.g.,][]{mkato1994, hachisu2003},
    or from the optically thick gas of the rotating accretion disk 
    with a Keplerian velocity of 1500 $\mathrm{km\ s}^{-1}$ at the inner disk edge \citep{mason2012}.
    Here, the narrow component refers to a relatively narrow emission profile compared to much broader components, 
    such as the initial nova ejecta with a full width at zero intensity (FWZI) velocity of approximately 10000 $\mathrm{km\ s}^{-1}$ \citep{anupama2013}.
    We also present a quiescent spectrum of U Sco observed in 2024 as a comparison with the spectra during the 2022 eruption.
    In section \ref{sec02:obs}, we describe our optical spectroscopy of U Sco.
    Our results on the narrow emission line components are summarized in section \ref{sec03:res}.
    Discussion and summary follow in sections \ref{sec04:dis} and \ref{sec05:sum}, respectively.
   
\section{Observations}\label{sec02:obs}

\begin{figure}[tb]
        \begin{center}
            \includegraphics[width=8cm]{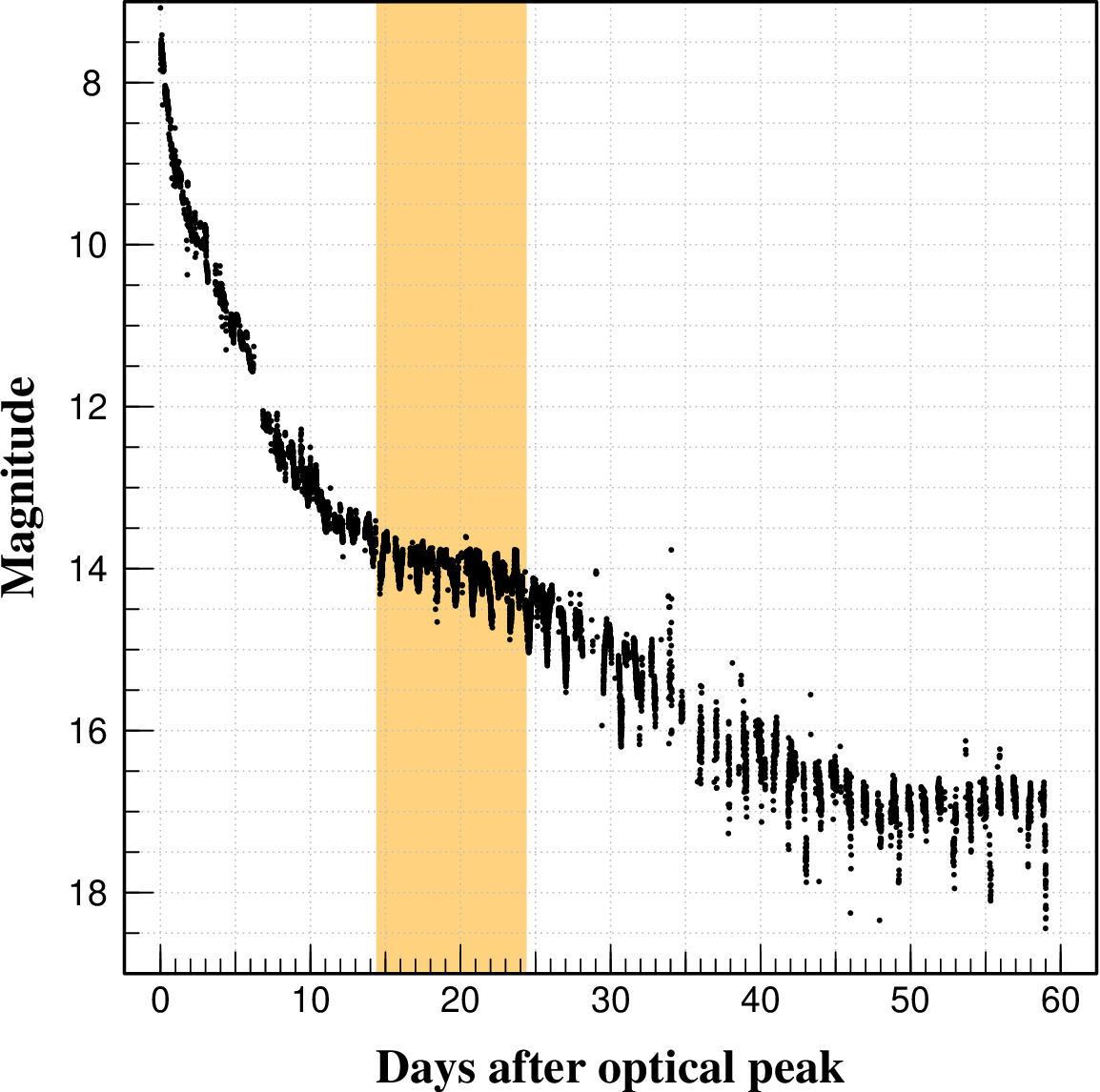} 
        \end{center}
        \caption{Optical light curve of the U Sco 2022 nova eruption (black filled circles).
            The orange region represents the optical plateau stage roughly between days 14 and 24 after the optical peak.
            {Alt text: Scatter plot for the light curve.  
            The x-axis represents days after the optical peak of the U Sco 2022 eruption. 
            The y-axis represents the optical magnitude.} 
        }\label{fig:lc2022}
\end{figure}

    In this section, we describe our optical spectroscopy of U Sco during the 2022 nova eruption and the 2024 quiescent phase.
    It should be noted that \citet{rudy2023} present infrared and optical spectra on days 2.54, 3.60, and 8.59 after the U Sco 2022 eruption, 
    and that \citet{evans2023} provide a detailed summary of near-infrared spectra between days 5.4 and 45.6.
    For reference, figure \ref{fig:lc2022} shows the optical light curve of the U Sco 2022 eruption (black circles), excerpted from \citet{muraoka2024}.
    These photometric data and our following spectroscopic data are available via the DOI links provided in the data availability section.

\subsection{The 2022 nova eruption}

    \begin{table*}[tb]
        \tbl{List of observers and instruments for spectroscopic observations during the U Sco 2022 eruption.}{%
        \begin{tabular}{cccc}
            \hline
                Code & Observer (Observatory) & Instrument & Resolution $(\lambda/\Delta\lambda)$ \\ 
            \hline
                ALE & Arthur Leduc & 20 cm F/5 Newton telescope, Star'Ex BR 23 microns, and ASI183MM Pro CMOS & 600  \\
                BHMA & Hamish Barker & 25 cm Newton telescope, L200 Littrow Spectrograph & 500  \\ 
                BHQ & Terry Bohlsen & C11 F/6.4 telescope, Shelyak LISA, and ASI183MM Pro CMOS & 1500  \\
                BRAJ & Raul Bruzzone & 35 cm F/10 telescope, Shelyak Alpy600, and Atik414ex CCD & 600  \\ 
                FAS & Forrest Sims & 50 cm F/6.8 PlaneWave CDK20 telescope, Shelyak LISA, and Atik414ex CCD & 1000 \\
                FJQ & James Foster & 43 cm F/6.8 CDK17 telescope, Shelyak LISA, in IR mode and Atik460ex CCD & 750  \\
                 & & Shelyak Alpy600 and Atik414ex CCD & 500 \\
                FKO & Mitsugu Fujii & 40 cm F/10 telescope, FBSPEC-V, and ASI2600MM Pro CMOS & 500  \\
                 & (Fujii Kurosaki Obs.\footnotemark[$*$]) & & \\
                KSH & Keith Shank & C14 F/7.6 telescope, Shelyak LISA, and SX825 Pro CCD & 1000 \\
                PAD & Pavol A. Dubovsky & 14" Schmidt-Cassegrain telescope, Shelyak LISA, and Atik460ex CCD & 1000  \\
                PCA & Paolo Cazzato & 25 cm F/4.8 VX10 telescope, Shelyak Alpy600, and Atik414ex CCD & 530  \\
                2SPOT & Southern Spectroscopic Project  & 30.5 cm F/4 Newton telescope, Shelyak Alpy600, and Atik414ex CCD & 600   \\
                 & Observatory Team & & \\
            \hline
        \end{tabular}}\label{tab02:obs2022}
        \begin{tabnote}
            \footnotemark[$*$] The spectra gallery is available at <https://otobs.org/FBO/index.html>. \\
        \end{tabnote}
    \end{table*}

    \begin{table}[tb]
        \tbl{Log of spectroscopic observations during the U Sco 2022 eruption.\footnotemark[$*$]}{%
        \begin{tabular}{rrccc}
            \hline
                mid BJD(TT) & $\Delta t$\footnotemark[$\dagger$] & Orbital phase\footnotemark[$\ddagger$]& Code\footnotemark[$\S$] & Database \\
                $(2400000+)$ & (day) & & &\\ 
            \hline
                59737.764 & 0.09 & 0.12 & FAS & BAA, ARAS \\ 
                59737.989 & 0.31 & 0.30 & FKO & - \\
                59738.866 & 1.19 & 0.01 & FJQ & BAA, AVSpec \\
                59738.968 & 1.29 & 0.10 & BHQ & AVSpec \\ 
                59739.940 & 2.26 & 0.89 & BHMA & AVSpec \\
            \hline
        \end{tabular}}\label{tab02:log2022}
        \begin{tabnote}
            \footnotemark[$*$] This is just a sample of e-table 1. \\
            \footnotemark[$\dagger$] Days after the optical peak of the U Sco 2022 eruption. \\
            \footnotemark[$\ddagger$] Calculated by equation (1) of \citet{muraoka2024}. \\
            \footnotemark[$\S$] Observer's code (see table \ref{tab02:obs2022}). \\
        \end{tabnote}
    \end{table}

    The U Sco 2022 nova eruption was firstly reported by Masayuki Moriyama,
    a member of the Variable Star Observers League in Japan (VSOLJ),\footnote{<http://vsolj.cetus-net.org/index.html>.}
    on 2022 June 6.72 UTC [BJD(TT)\footnote{Barycentric Julian date based on the terrestrial time.} 2459737.23; 
    vsnet-alert 26798\footnote{<http://ooruri.kusastro.kyoto-u.ac.jp/mailarchive/vsnet-alert/26798>.}].
    The time of the maximum optical brightness during the 2022 eruption was BJD(TT) 2459737.68(3) \citep{muraoka2024}. 
    From the time of this peak, we obtained 62 optical spectra over a period of $\sim$22 d.
    Observers (observatories) and instruments are summarized in table \ref{tab02:obs2022}.
    The log of this spectroscopy is summarized in 
    table \ref{tab02:log2022}.\footnote{A complete listing of table \ref{tab02:log2022} is available as e-table 1 in the supplementary data section.}
    Data reduction was performed by each observer.
    Most of the data are also available from the following archival databases: 
    British Astronomical Association (BAA),\footnote{<https://britastro.org/specdb/data.php>.}
    American Association of Variable Star Observers (AAVSO) Spectroscopic Database (AVSpec),\footnote{<https://apps.aavso.org/avspec/search>.}
    and Astronomical Ring for Amateur Spectroscopy \citep[ARAS;][]{ARAS} Spectral Database.\footnote{<https://aras-database.github.io/database/index.html>.}

\subsection{The 2024 quiescent phase}

    To obtain a quiescent spectrum of U Sco after the 2022 eruption, 
    we performed a spectroscopic observation in optical using the fiber-fed integral field spectrograph \citep[KOOLS-IFU;][]{KIF}
    mounted on the 3.8-m Seimei telescope \citep{seimei}.
    We used EB656 as a grism, which provides a resolution of $R \sim 2000$ and covers a wavelength range of  5700--7800 \AA.
    The midpoint of our observation was BJD(TT) 2460534.017, 794 d after the optical peak of the 2022 eruption.\footnote{
    The optical magnitude was 17.5 on BJD(TT) 2460534 from the ASAS-SN Sky Patrol V2.0 \citep{ASN_3,ASN_1}.}
    The orbital phase was 0.19, calculated by equation (1) of \citet{muraoka2024} [see also equation (1) of \citealp{schaefer2011}, which is described in HJD].
    We captured 6 frames, each with an exposure time of 900 s.
    Data reduction was performed using IRAF in the standard manner 
    (bias subtraction, flat fielding, wavelength calibration with arc lamps, aperture determination, spectral extraction, 
    and flux calibration with the standard star HD147283).
 
\section{Results}\label{sec03:res}
    
\subsection{Overall spectral evolution}\label{subsec:2022spec}

    \begin{figure}[tb]
        \begin{center}
            \includegraphics[width=8cm]{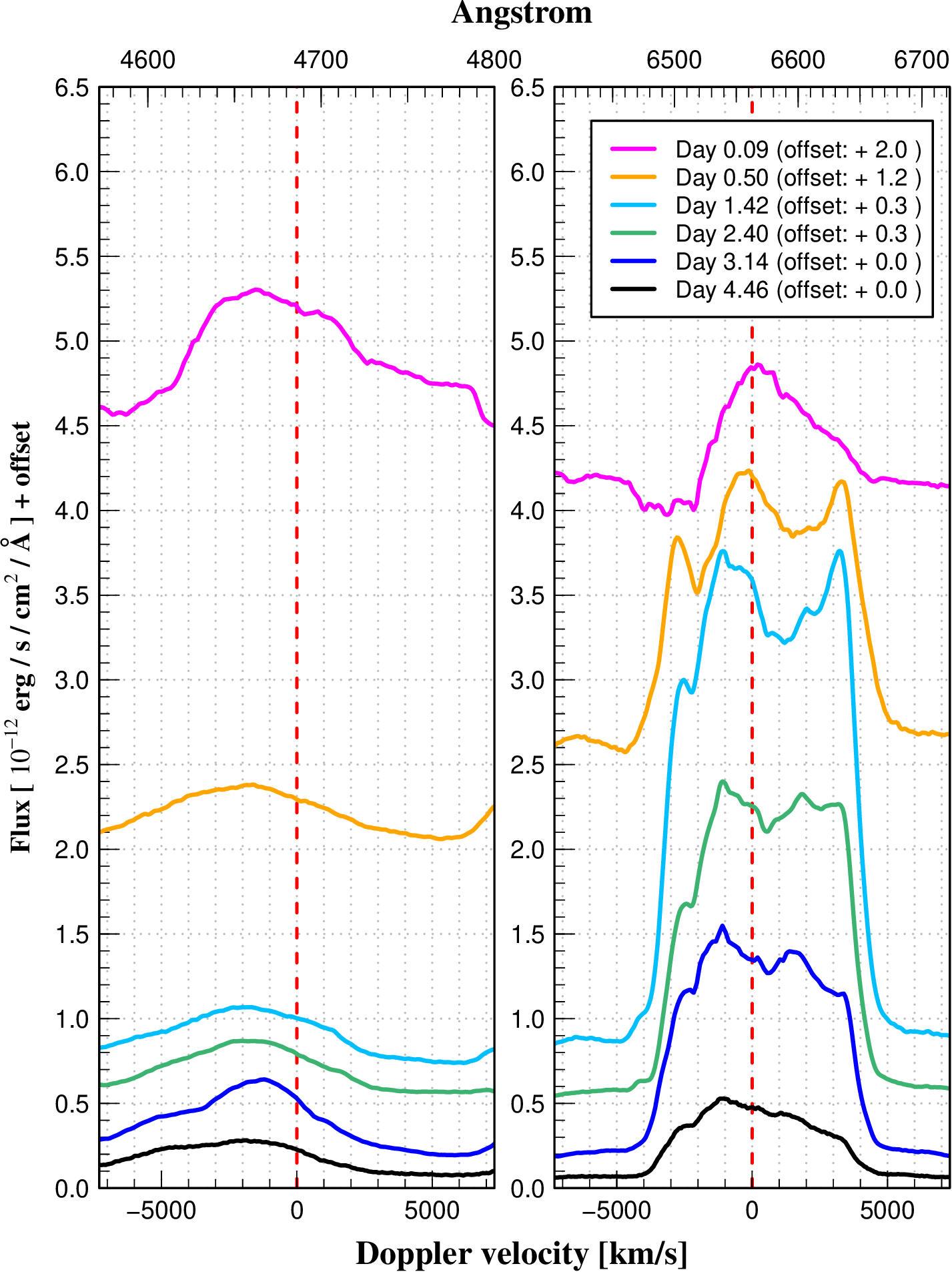} 
        \end{center}
        \caption{Selected line profiles of He \textsc{ii} 4686 (left panel) and H$\alpha$ (right panel) from flux-calibrated spectra
        during days 0.09--4.46 after the optical peak of the U Sco 2022 eruption. 
        The red dashed lines on the left and right represent the rest wavelength positions of He \textsc{ii} 4686 and H$\alpha$, respectively.  
            {Alt text: Two panels, each containing 6 line profiles. 
            The upper x-axes in both panels represent wavelength in units of Angstrom,
            and the lower x-axes in both panels represent the corresponding Doppler velocity in units of kilometer per second.
            The y-axes represent flux density in units of 10 to the power of minus 12 erg per second per square centimeter per Angstrom plus offset.} 
        }\label{fig:spec_1}
    \end{figure}

    \begin{figure}[tb]
        \begin{center}
            \includegraphics[width=8cm]{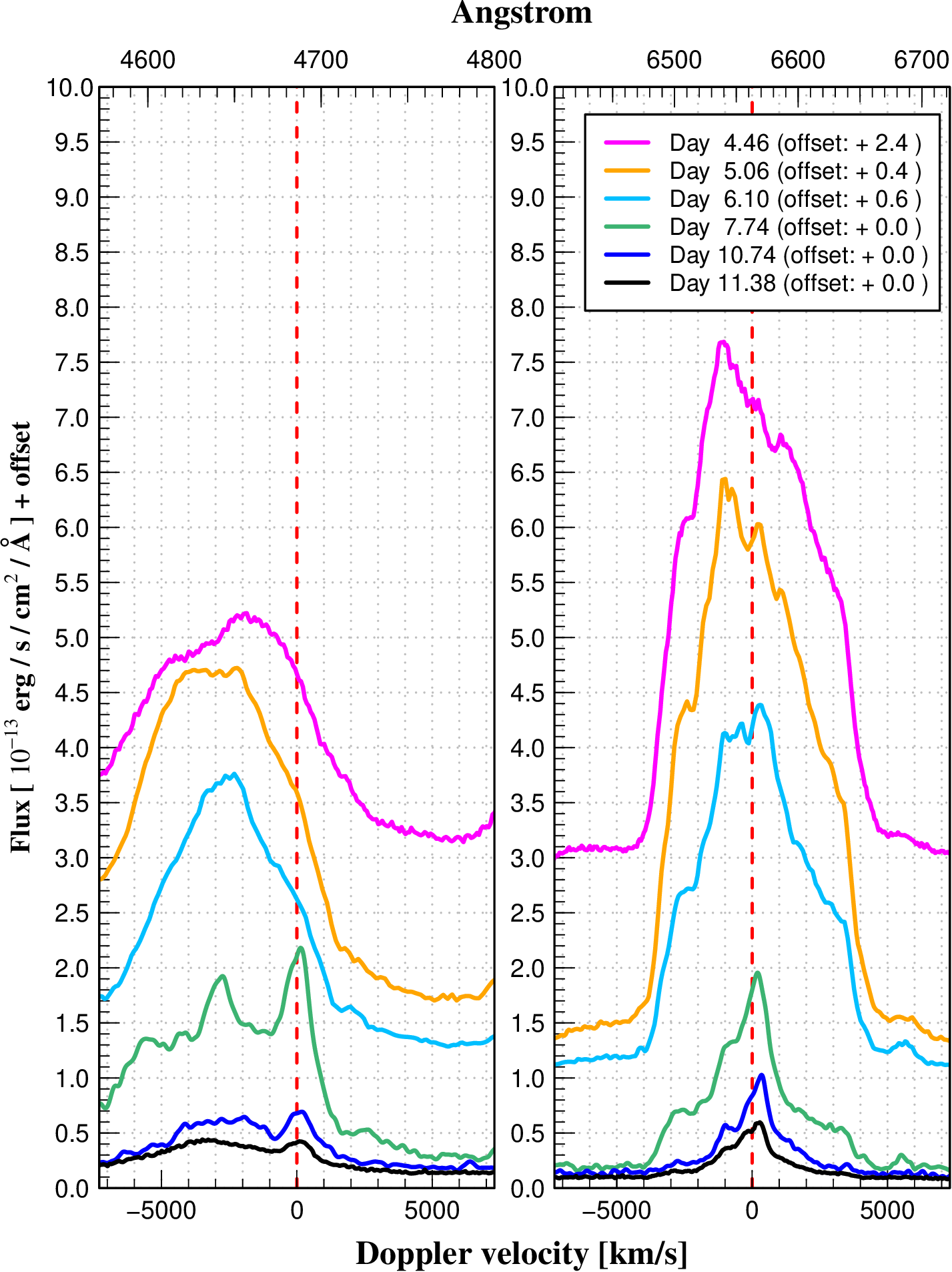} 
        \end{center}
        \caption{Selected line profiles of He \textsc{ii} 4686 (left panel) and H$\alpha$ (right panel) during days 4.46--11.38.    
            {Alt text: Two panels, each containing 6 line profiles. 
            The upper x-axes in both panels represent wavelength in units of Angstrom,
            and the lower x-axes in both panels represent the corresponding Doppler velocity in units of kilometer per second.
            The y-axes represent flux density in units of 10 to the power of minus 13 erg per second per square centimeter per Angstrom plus offset.} 
        }\label{fig:spec_2}
    \end{figure}

    \begin{figure}[tb]
        \begin{center}
            \includegraphics[width=8cm]{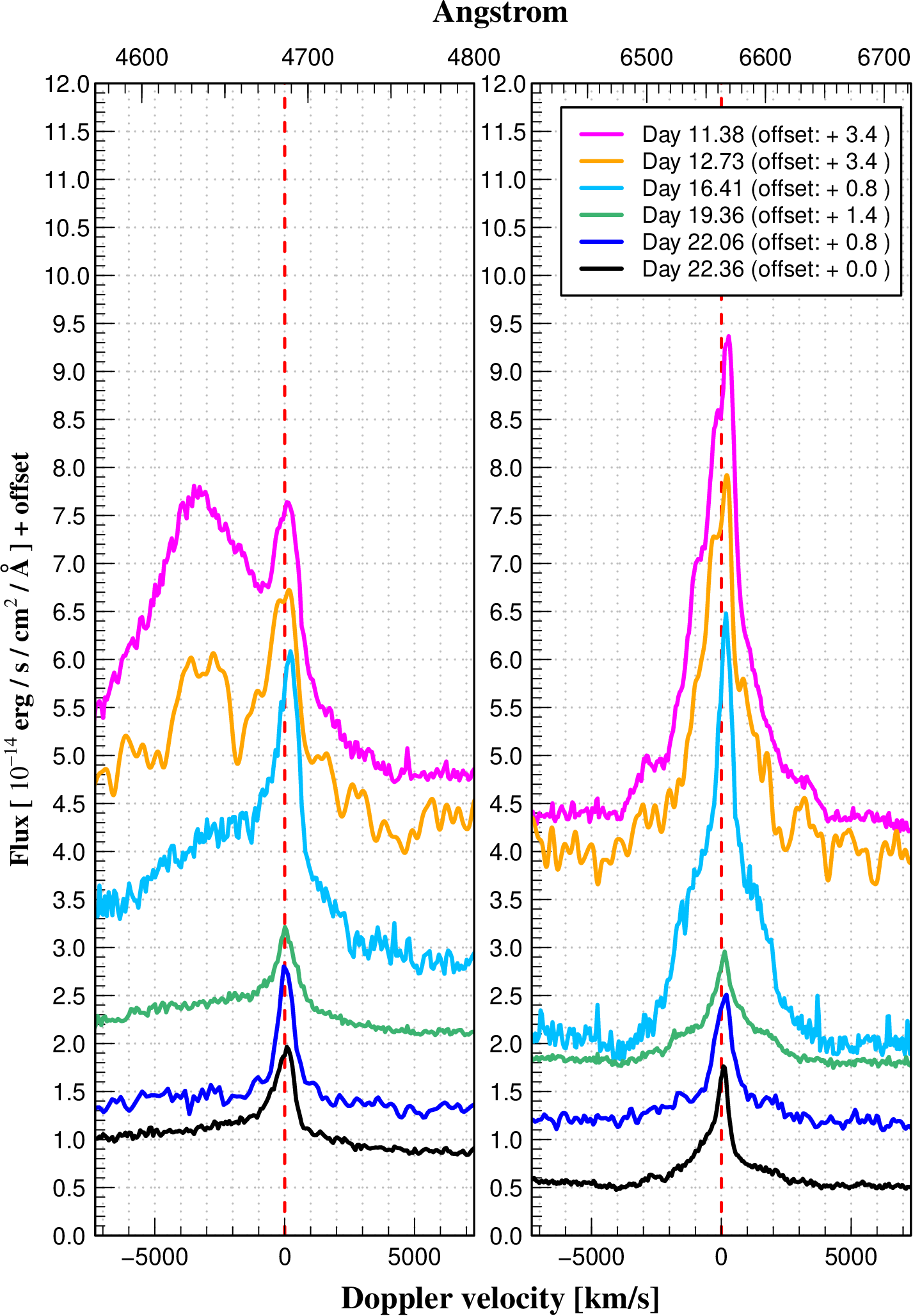} 
        \end{center}
        \caption{Selected line profiles of He \textsc{ii} 4686 (left panel) and H$\alpha$ (right panel) during days 11.38--22.36.    
            {Alt text: Two panels, each containing 6 line profiles. 
            The upper x-axes in both panels represent wavelength in units of Angstrom,
            and the lower x-axes in both panels represent the corresponding Doppler velocity in units of kilometer per second.
            and the y-axes represent flux density in units of 10 to the power of minus 14 erg per second per square centimeter per Angstrom plus offset.} 
        }\label{fig:spec_3}
    \end{figure}

    Figures \ref{fig:spec_1}--\ref{fig:spec_3} show the overall spectral evolution of 
    H$\alpha$ and He \textsc{ii} 4686 line profiles during the U Sco 2022 eruption.
    These spectra are excerpts from those which have been flux-calibrated.
    We also present our all optical spectra in e-figures 1--4.\footnote{
    \ E-figures 1--4 are available in the supplementary data section.}
    
    On day 0.09 after the optical peak, a P Cyg profile was observed in the H$\alpha$ line profile, 
    showing blue-side absorption with a velocity of 2000--4000 $\mathrm{km\ s}^{-1}$
    (magenta line in the right panel of figure \ref{fig:spec_1}).
    However, this feature was barely visible on day 0.50 (orange line in the right panel of figure \ref{fig:spec_1}).
    After day $\sim$1, this P Cyg profile was completely lost, 
    and a boxy broad component of H$\alpha$ was clearly detected in figure \ref{fig:spec_1},
    with a velocity of 4000--5000 $\mathrm{km\ s}^{-1}$ in its tail.
    On the other hand, the whole emission line of He \textsc{ii} 4686 did not yet appear in the left panel of figure \ref{fig:spec_1},
    and it was dominated by the Bowen blend \citep{BOWEN}.
    The Bowen emission around He \textsc{ii} 4686 mainly consists of C \textsc{iii}, C \textsc{iv}, N \textsc{iii} and O \textsc{ii} \citep{schmidtobreick2003}.

    Figure \ref{fig:spec_2} shows the spectra from the period when the optical light curve was in the early decline stage
    (see also figure \ref{fig:lc2022}).
    The narrow components of both H$\alpha$ and He \textsc{ii} 4686 did not appear until day $\sim$6.
    However, the narrow component of He \textsc{ii} was faintly detected on days 6.80 and 7.01, as presented in e-figure 3. 
    On day 7.74 (green line in the left panel of figure \ref{fig:spec_2}), 
    it was clearly visible and its flux became comparable to that of the Bowen blend.
    Around the same time, the narrow component of H$\alpha$ developed, 
    and a single-peaked profile was clearly detected within $\pm$1000 $\mathrm{km\ s}^{-1}$ of the rest wavelength of H$\alpha$
    in the right panel of figure \ref{fig:spec_2}.
    
    Figure \ref{fig:spec_3} shows the spectra from the period when the optical light curve was mostly in the optical plateau stage,
    roughly between days 14 and 24 (\citealp{muraoka2024}; see also the orange region in figure \ref{fig:lc2022}).\footnote{
    \ The "second" optical plateau stage was observed after day $\sim$50. 
    It was also detected during the 2009 nova eruption of KT Eri 
    and is considered to result from a viscous heating disk together with a high mass-transfer rate \citep{hachisu2025}.}
    Just before the onset of the optical plateau stage, a peak was also observed in the Bowen blend 
    (magenta and orange lines in the left panel of figure \ref{fig:spec_3}).
    However, as the optical plateau stage began, this peak became less prominent, 
    and the single-peaked narrow component of He \textsc{ii} 4686 became more pronounced.
    Similarly, the H$\alpha$ line showed the single-peaked narrow component that emerged prominently 
    throughout the optical plateau stage (right panel of figure \ref{fig:spec_3}).

    In our observations, we did not clearly detect the broad component of He \textsc{ii} 4686.
    The narrow component of H$\beta$ also showed a similar behavior to that of H$\alpha$, as presented in e-figures 1--4.
    However, we did not clearly detect the narrow components of other lines, such as He \textsc{i}.

\subsection{Evolution of line width during the 2022 eruption}\label{subsec:2022lw}
 
    \begin{figure*}[tb]
        \begin{center}
            \includegraphics[width=\linewidth]{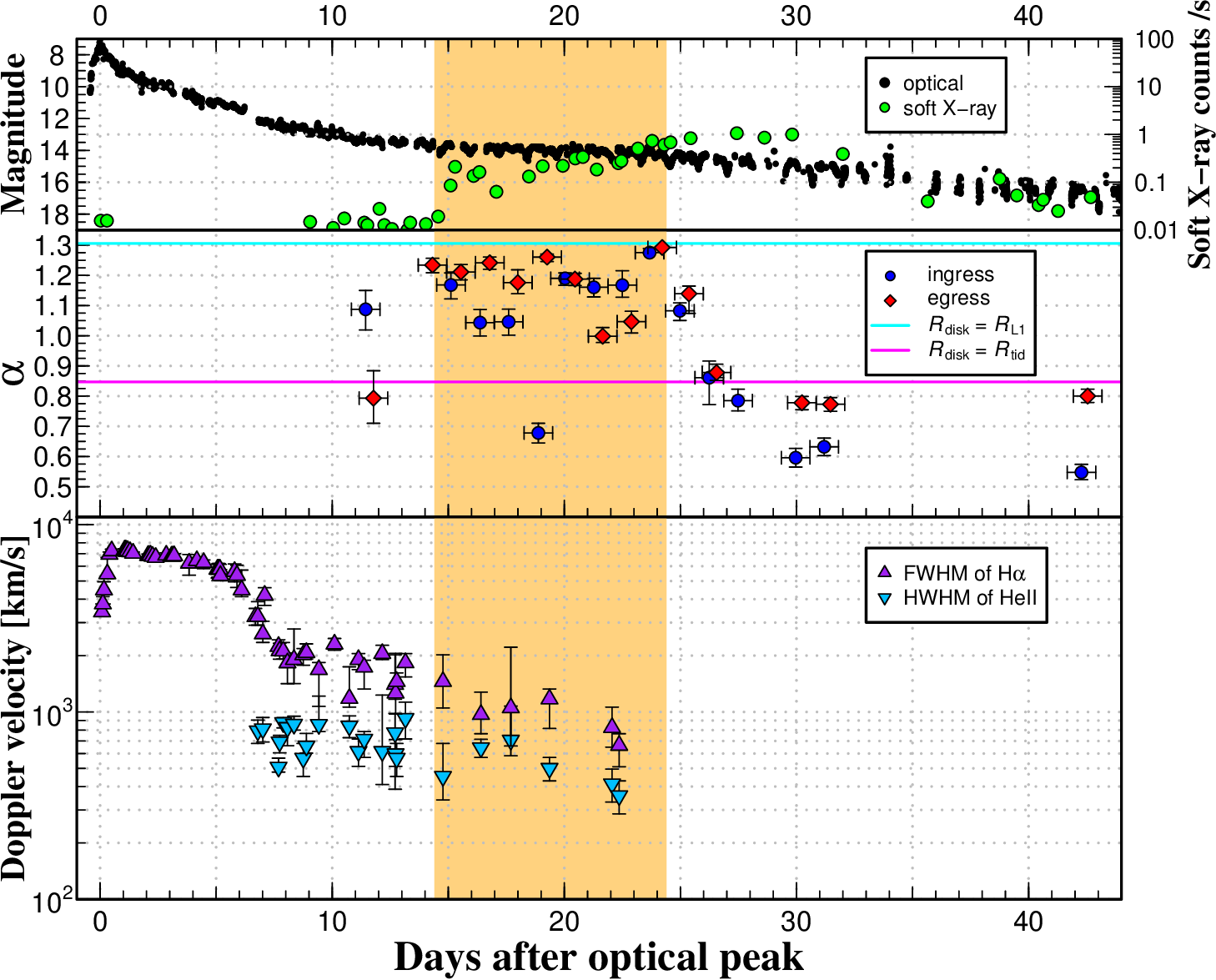} 
        \end{center}
        \caption{\textbf{Upper panel}: Optical (black filled circles) and soft X-ray (green filled circles) light curves of the U Sco 2022 eruption,
            excerpted from the upper panel of figure 5 in \citet{muraoka2024}.
            The orange region represents the optical plateau stage.
            \textbf{Middle panel}: Transition of $\alpha$, the normalized outer radius of the optical light source centered at the WD,
            estimated by the ingress (blue filled circles) and egress (red filled diamonds) phases of the primary optical eclipse.
            This is the result of reanalyzing \citet{muraoka2024}.
            The cyan line represents the disk radius when expanding to the distance between the primary WD and the L1 point $R_{\mathrm{L1}}$.
            The magenta line represents the disk radius when expanding to the tidal truncation radius $R_{\mathrm{tid}}$.
            \textbf{Lower panel}: Transition of the FWHM of H$\alpha$ (purple filled triangles) 
            and the HWHM of He \textsc{ii} 4686 (sky-blue filled inverted triangles).
            {Alt text: Three scatter plots.  
            The x-axes in all panels represent days after the optical peak of the U Sco 2022 eruption. 
            In the upper panel, the left y-axis represents the optical magnitude, 
            and the right one represents the soft X-ray counts per second on a logarithmic scale.
            The y-axis in the lower panel represents the line width in terms of Doppler velocity 
            in units of kilometer per second on a logarithmic scale.} 
        }\label{fig:width}
    \end{figure*}

    Figure \ref{fig:width} shows the evolution of the emission line width, 
    combined with the results of the photometric analysis from \citet{muraoka2024}.
    The upper panel of figure \ref{fig:width} shows the optical (black circles) and soft X-ray (green circles) light curves of the U Sco 2022 eruption.
    The orange region represents the optical plateau stage.
    The middle panel of figure \ref{fig:width} shows the transition of $\alpha$. 
    During and after the optical plateau stage,
    the estimated outer disk radius $R_\mathrm{disk}$ can be written as $R_{\mathrm{disk}} = \alpha R_1^{*}$
    (see subsection 4.2 of \citealp{muraoka2024} for a detailed discussion),
    where $R_1^{*}$ represents the primary's Roche volume radius
    (e.g., section 4.4 of \citealp{frank2002}; \citealp{leahy2015}; see figure 7 of \citealp{muraoka2024} for the Roche geometry of U Sco).\footnote{
    \ It is an indicator of the average radius of the non-spherical Roche lobe.
    A sphere of radius $R_1^{*}$ has the same volume as the region surrounded by the primary’s Roche lobe.
    We adopt an approximate equation proposed by \citet{eggleton1983} [see also equation (A4) of \citealp{muraoka2024}].}
    It should be noted that this is the result of reanalyzing \citet{muraoka2024}, 
    using the Markov chain Monte Carlo (MCMC) method with the No-U-Turn Sampler \citep[NUTS;][]{NUTS}
    to measure the ingress and egress phases of the primary optical eclipse,
    as implemented in the Python module PyMC version 5 \citep{PyMC5}.\footnote{
    \ <https://www.pymc.io/welcome.html>.
    }
    We used this modified method to obtain an accurate 99\% confidence interval as the error estimate.
    We confirmed convergence of the iterative Monte Carlo simulations 
    by ensuring that the potential scale reduction $\hat{R} < 1.1$ was satisfied
    \citep[e.g., subsection 11.5 of][]{PSR}.
    The lower panel of figure \ref{fig:width} shows the transition of 
    the full width at half maximum (FWHM) of H$\alpha$ (purple triangles) and
    the half width at half maximum (HWHM) of He \textsc{ii} 4686 (sky-blue inverted triangles),
    expressed in Doppler velocity units with respect to their corresponding rest wavelength positions.
    
    The FWHM of H$\alpha$ was roughly estimated as follows.
    First, the continuum surrounding the emission line was modeled using a linear fitting, 
    excluding the wavelength region within $\pm$150 \AA\ of the rest wavelength of H$\alpha$.
    The emission line was then extracted by subtracting this fitted continuum from the original spectrum.
    The peak flux of the emission line was identified, and the two wavelength positions corresponding to half of its peak flux were determined.
    Finally, we calculated the difference between these two as the FWHM.

    For the HWHM of He \textsc{ii} 4686, the continuum surrounding the emission line was estimated
    by assuming a constant flux level in the wavelength region $\sim$100 \AA\ redward of the rest wavelength of He \textsc{ii}.
    Then, we similarly subtracted this assumed continuum from the original spectrum to extract the emission line.
    Since this extracted profile also includes the Bowen blend on the blue side, 
    we measured only the redward wavelength position corresponding to half of its peak flux as the HWHM.
    It should be noted that the extracted He \textsc{ii} line profile 
    may be contaminated by the broad component of the Bowen blend in its tail,
    so we may overestimate the HWHM of He \textsc{ii} in our method.
    
    The FWHM of H$\alpha$ was approximately 3500 $\mathrm{km\ s}^{-1}$ on day 0.1 after the optical peak of the U Sco 2022 eruption,
    but it rapidly increased to around 7000 $\mathrm{km\ s}^{-1}$ by day 0.5.
    It then decreased slightly on a logarithmic scale until day $\sim$6, when it reached approximately 5000 $\mathrm{km\ s}^{-1}$.
    Subsequently, between roughly days 6 and 8, the FWHM of H$\alpha$ decreased sharply, reaching around 2000 $\mathrm{km\ s}^{-1}$.
    In the middle of the optical plateau stage, it reached a value of approximately 1000 $\mathrm{km\ s}^{-1}$, 
    and in the latter part, it further decreased to around 700 $\mathrm{km\ s}^{-1}$.

    Meanwhile, the HWHM of He \textsc{ii} 4686 is measured only from day 6.80 and later
    because its narrow component became distinguishable from the Bowen blend starting on that day.
    Before the optical light curve entered the optical plateau stage, the HWHM of He \textsc{ii} was approximately 700 $\mathrm{km\ s}^{-1}$.
    In the latter part of the optical plateau stage, it decreased to around 400 $\mathrm{km\ s}^{-1}$.

\subsection{Quiescent spectrum in 2024}\label{subsec:quie}

    \begin{figure}[tb]
        \begin{center}
            \includegraphics[width=8cm]{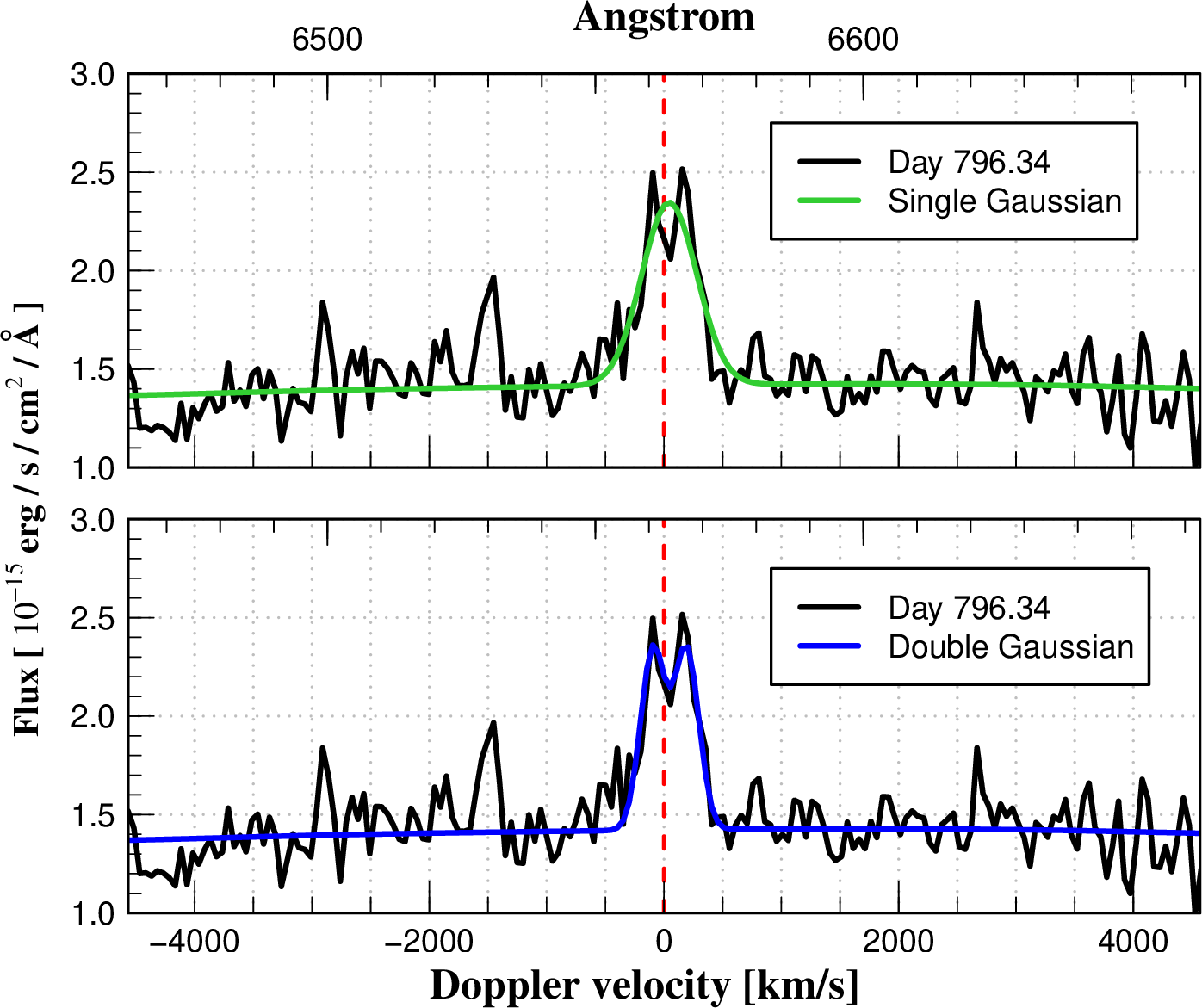} 
        \end{center}
        \caption{H$\alpha$ line profile of U Sco during the 2024 quiescent phase (black solid lines).    
            The green line in the upper panel represents the single-Gaussian fit.
            The blue line in the lower panel represents the double-Gaussian fit.
            {Alt text: Two line graphs showing the fitting results of the H$\alpha$ line profile. 
            The upper x-axes in both graphs represent wavelength in units of Angstrom,
            and the lower x-axes in both graphs represent the corresponding Doppler velocity in units of kilometer per second.
            The y-axes represent flux density in units of 10 to the power of minus 15 erg per second per square centimeter per Angstrom.} 
        }\label{fig:spec_quie}
    \end{figure}

    Figure \ref{fig:spec_quie} shows the quiescent spectrum approximately 2 yr after the U Sco 2022 eruption (black lines).
    We only obtained the H$\alpha$ line profile due to the wavelength range of EB656 grism.
    Unlike the single-peaked profile observed during the 2022 eruption, 
    we observed a double-peaked profile, likely originating from the accretion disk.
    To measure the FWHM and the double-peak separation (DP), 
    we extracted the H$\alpha$ line profile from the continuum and performed the following two fittings.

    The first fitting was a single-Gaussian fitting for measuring the FWHM.
    We followed the method of \citet{casares2015}, 
    whose fitted model consists of a constant plus a Gaussian function in the wavelength region within $\pm$200 \AA\ of the rest wavelength of H$\alpha$.
    Namely, the flux density of H$\alpha$, $y$, can be modeled as the following function of wavelength or corresponding Doppler velocity, $x$:
    \begin{eqnarray}
        y = A \exp \left[ -\frac{(x-\mu)^2}{2\sigma^2} \right] + C .     
    \end{eqnarray}  
    \noindent
    The fitting parameters are $A$, $\mu$, $\sigma$, and $C$.
    The FWHM can be derived as $\mathrm{FWHM} = 2\sqrt{2\ln{2}}\sigma$.
    The green line in the upper panel of figure \ref{fig:spec_quie} represents the result of this single-Gaussian model.

    The second fitting was a double-Gaussian fitting for measuring the DP.
    We followed the method of \citet{casares2016}, 
    whose fitted model consists of a constant plus two Gaussian functions of identical amplitude and standard deviation  
    in the wavelength region within $\pm$200 \AA\ of the rest wavelength of H$\alpha$.
    Namely, the flux density of H$\alpha$, $y$, can be modeled as the following function of wavelength or corresponding Doppler velocity, $x$:
    \begin{eqnarray}
         y = A \left\{ \exp \left[ -\frac{(x-\mu_1)^2}{2\sigma^2} \right] + \exp \left[ -\frac{(x-\mu_2)^2}{2\sigma^2} \right] \right\} + C .
    \end{eqnarray} 
    \noindent
    The fitting parameters are $A$, $\mu_1$, $\mu_2$, $\sigma$, and $C$.
    The DP can be derived as $\mathrm{DP} = |\mu_1 - \mu_2|$.    
    The blue line in the lower panel of figure \ref{fig:spec_quie} represents the result of this double-Gaussian model.

    These fitting parameters were estimated using the MCMC method with NUTS,
    and the errors were derived based on the 99\% confidence interval.
    We estimated that the FWHM to be 552(9) $\mathrm{km\ s}^{-1}$  
    and the DP to be 287(3) $\mathrm{km\ s}^{-1}$. 
    
\section{Discussion}\label{sec04:dis}

\subsection{Comparison with previous U Sco eruptions}

    During the U Sco 2022 eruption, the H$\alpha$ line profile showed a narrow single peak
    on day 16.41 after the optical peak (sky-blue line in the right panel of figure \ref{fig:spec_3}).
    On day +17 during the U Sco 2010 eruption, a similar feature was observed in the H$\alpha$ line profile \citep{mason2012},
    whereas a triple-peaked profile was observed on day 17 during the U Sco 1999 eruption \citep{iijima2002}.
    The radial velocities of these triple peaks were 
    $-$1560 $\mathrm{km\ s}^{-1}$, $+$60 $\mathrm{km\ s}^{-1}$, and $+$1800 $\mathrm{km\ s}^{-1}$ \citep{iijima2002},
    and their peak fluxes were approximately one-fourth of the peak flux of H$\alpha$ on day 16.41 during the 2022 eruption.
    While the central peak of H$\alpha$ showed nearly the same velocity among these three nova eruptions,
    the 1999 eruption additionally displayed two peaks on both the red and blue sides, 
    which were located within the wings of the H$\alpha$ profiles in 2010 and 2022.

    \citet{hachisu2024} have stated that, after the optical peak, 
    the nova wind ejected later catches up with the one ejected earlier, creating a reverse shock. 
    As a result, a large part of the nova ejecta becomes confined to the shocked shell, 
    leading to a main source of broad emission lines even in the SSS stage 
    with a velocity of approximately 2000 $\mathrm{km\ s}^{-1}$ \citep{hachisu2025}.
    \citet{yamanaka2010} have noted that the apparent geometry or shape of the ejecta,
    which depends on the orbital phase at the onset of the nova eruption, could influence the observed line profiles.
    The orbital phases at optical peaks are approximately 0.41 for the 1999 eruption \citep{yamanaka2010},
    0.05 for the 2010 eruption \citep{yamanaka2010}, and 0.05 for the 2022 eruption \citep[based on][]{muraoka2024}.
    Therefore, assuming that it takes about half an orbital phase for the nova eruption to reach the optical peak after it occurs,
    the secondary was likely located between the observer and the primary WD at the onset of the 1999 eruption.
    This might alter the mass distribution of the ejecta or shocked shell on the observer's side, 
    resulting in the different line profiles from those in 2010 and 2022. 
    
    In another case, RS Oph\footnote{
    \ The orbital period is 453.6(4) d \citep{brandi2009}.}
    recently experienced two different nova eruptions in 2006 and 2021.
    \citet{azzollini2023} have studied the optical spectra during both eruptions 
    and suggested that there was no significant difference between their line profiles.
    \citet{page2022_3} have pointed out that the orbital phases at the onset of the 2006 and 2021 eruptions were 0.26 and 0.72, respectively.
    Therefore, the secondary was off the line of sight between the primary WD and the observer during both eruptions,
    which may have not affected the mass distribution of the ejecta or shocked shell on the observer's side,
    resulting in the similar line profiles.

    Regarding the optical quiescent spectrum, 
    \citet{maxwell2014} present some quiescent spectra in 2011 on days 445 and 460 after the U Sco 2010 eruption.
    In these 2011 spectra, a double-peaked profile was observed near the rest wavelength of H$\alpha$, 
    similar to that seen in the 2024 quiescent spectrum (black lines in figure \ref{fig:spec_quie}).
    In addition, on both the red and blue sides within $\pm$100 \AA\ of the rest wavelength of H$\alpha$,
    there were some peaks with fluxes comparable to that of the double-peaked profile in the 2011 quiescent spectra,
    though these features were only weakly detectable in the 2024 quiescent spectrum.
    We consider that these originated from components other than the accretion disk, 
    possibly from the circumstellar material formed by ejecta.
    The 2024 spectrum was taken on day 794 after the 2022 eruption, 
    which is around 1 yr later than the 2011 spectra taken on days 445 and 460 after the 2010 eruption, 
    so we suspect that the contribution from the circumstellar material might have weakened compared to the disk component in 2024.

    \citet{schaefer2022} has also mentioned that the orbital period changes across various U Sco nova eruptions. 
    Since these observational differences may be associated with such orbital period changes,
    it is necessary to continue monitoring future U Sco nova eruptions and to compare them with previous ones.

\subsection{Origin of the narrow components}

    As mentioned in subsections \ref{subsec:2022spec} and \ref{subsec:2022lw}, 
    the narrow component of He \textsc{ii} 4686 was faintly detected on days 6.80 and 7.01,
    and it became clearly visible on day 7.74 after the optical peak of the U Sco 2022 eruption.
    Simultaneously, between roughly days 6 and 8, the FWHM of H$\alpha$ decreased sharply, 
    suggesting the development of the narrow component of H$\alpha$.
    In our observations, there was no significant difference when the narrow components of these lines began to appear.
    \citet{muraoka2024} have reported that the primary optical eclipse became detectable from 10.4--11.6 d after the optical peak
    (see also the middle panel of figure \ref{fig:width}). 
    Therefore, it is found that the narrow components of H$\alpha$ and He \textsc{ii} 4686 appeared 
    earlier than the onset of the optical eclipses of the accretion disk.
    This is consistent with the result of \citet{anupama2013}, 
    suggesting that the He \textsc{ii} 4686 line appeared just before the onset of the optical plateau stage,
    weakly present on day $\sim$7 during the U Sco 2010 eruption.
    They have also stated that the narrow components of hydrogen and He \textsc{ii} could both arise in the same region.

    Regarding the origin of such narrow components,
    which were observed continuously even before the disk eclipses appeared,
    \citet{yamanaka2010} have proposed that the narrow component of H$\alpha$ on day $\sim$9 after the U Sco 2010 eruption 
    would be attributed to the nova wind near the WD.
    On the other hand, \citet{mason2012} have supported the hypothesis that 
    the narrow component is originating from the optically thick gas of the rotating accretion disk,
    indicating that it changed in position and profile with time and orbital phase during the U Sco 2010 eruption.
    However, \citet{anupama2013} did not find any indications of orbital flux variations in the He \textsc{ii} 4686 line during the same nova eruption.
    
    We support the hypothesis of the nova-wind origin.
    If these narrow components were mainly attributed to the optically thick disk, 
    the optical eclipses of the accretion disk should have been observable between roughly days 6 and 8.
    The absence of the disk eclipses during this period implies that the accretion disk was probably inside the ejecta photosphere. 
    It also suggests that the disk was not fully irradiated by the hotter WD and was not luminous enough to emit such clear narrow components in the optical range. 
    This may be because the mass-loss rate driven by the nova wind was still higher compared to the optical plateau stage, 
    causing a larger part of the energy flux emitted from the WD to be self-absorbed by the wind itself \citep[e.g.,][]{hachisu2006_2}.
    The theoretic light curve during the U Sco 1999 eruption, as modeled by \citet{hachisu2000}, also suggests that 
    the contribution of the accretion disk to the overall optical brightness was less dominant 
    compared to that of the ejecta photosphere before the onset of the optical plateau stage.

    \citet{mason2014} have also denied the origin of the accretion disk in the case of YY Dor, 
    arguing that the narrow component line forming region is closer to the WD than the broad one, 
    such as the inner most part of the ejecta.
    In our observations, the  H$\alpha$ and He \textsc{ii} 4686 line profiles did not show any significant changes or asymmetries related to the orbital phase
    (e.g., the blue line spectrum in figure \ref{fig:spec_3} for the orbital phase 0.97).
    Therefore, the line forming region of these narrow components is considered to be located at a distance on the order of the binary separation from the WD,
    closer to the binary system compared to the broad line forming region around the head of the nova ejecta.
    One can also see a detailed review of the narrow component in \citet{mason2014}.
    
    We were unable to track when the single peak of the narrow component disappeared during the U Sco 2022 eruption,
    but \citet{anupama2013} have reported that 
    the single-peaked narrow component of H$\alpha$ was still present on day $\sim$34 after the optical plateau stage
    and then lost on day $\sim$42 during the U Sco 2010 eruption.
    This coincided with the end of the soft X-ray stage on day $\sim$40, 
    when the soft X-ray count rate decreased to a tenth of its peak value \citep{muraoka2024}.
    We consider that the narrow components, mainly originating from the nova wind near the binary system, disappeared on that day
    because the steady H-burning on the WD surface weakened, leading to more diminished nova wind.

    \citet{hachisu2025} have also suggested that the outgoing disk surface flow, 
    which is a thin surface layer of the accretion disk blown in the nova wind,
    could be the origin of these narrow components.
    They have estimated that the disk surface flow is accelerated up to $\sim$1000 $\mathrm{km\ s}^{-1}$,
    which is consistent with our FWHM and HWHM results during the optical plateau stage. 
    The accretion disk became luminous in the optical range due to the irradiation by the emerging hotter WD during this period.
    As a result, the contribution of this disk surface flow, as well as that of the nova wind, might become evident, 
    making the narrow line profiles more pronounced.

    As mentioned in subsection \ref{subsec:quie}, 
    the FWHM of H$\alpha$ in quiescence was estimated to be 552(9) $\mathrm{km\ s}^{-1}$.
    Considering that the accretion disk was expanded due to the nova wind during the optical plateau stage, as demonstrated by \citet{muraoka2024},
    it is possible that the FWHM of the rotating accretion disk, apart from the outflowing disk matter, 
    might be smaller than the quiescent value during this period.
    Therefore, it is considered that the component of the nova wind overwhelmed that of the rotating accretion disk 
    in the H$\alpha$ line profile during the optical plateau stage.

\subsection{Interaction between the accretion disk and the nova wind}

    \citet{hachisu2003} have stated that a large velocity difference 
    between the surviving disk surface (several hundred $\mathrm{km\ s}^{-1}$) and the nova wind ($\sim$1000 $\mathrm{km\ s}^{-1}$) 
    drives a Kelvin-Helmholtz instability after the nova eruption.
    They have suggested that the internal density of the disk is much greater than that of the nova wind, 
    so only the very surface layer of the disk is dragged away like a free stream moving outward, 
    resulting in the expansion of the accretion disk.
    After the nova wind weakens or stops, the outer edge of the accretion disk shrinks toward the tidal truncation radius 
    in several dynamical timescales, namely, several orbital periods, 
    and finally the accretion disk settles into a steady state.

    Our observations align quite well with the model of \citet{hachisu2003},
    supporting that the nova wind even remains active several days after the nova eruption with a velocity of approximately 1000 $\mathrm{km\ s}^{-1}$,
    likely driving the expansion of the accretion disk until the end of the optical plateau stage.
    Based on this qualitative explanation, 
    our photometric and spectroscopic observations suggest the following scenario for the evolution of the U Sco 2022 nova eruption.\\

    \begin{enumerate}
    \renewcommand{\labelenumi}{\textbf{(\roman{enumi})}}
    \item 
    Just after the nova eruption occurred, the outer accreted layer expanded to initiate mass loss as optically thick nova wind, 
    and the ejecta photosphere centered at the primary WD rapidly expanded beyond the binary separation within one day.
    This initial mass loss exhibited a P Cyg profile in the emission line of H$\alpha$.
    Meanwhile, the accretion disk survived the nova eruption and was expanded close to the L1 point possibly due to the nova wind,
    but it was obscured in the photosphere immediately after the nova eruption.\\

    \item
    After the optical peak (day 0), the optical magnitude began to decrease because of the photosphere receding toward the WD.
    Due to the decrease in the ejecta density, the absorption along the line of sight weakened, and the P Cyg profile gradually disappeared.
    A boxy broad component, probably originating from the shocked shell formed by the initial ejecta, was clearly detected in the H$\alpha$ line profile. 
    The nova wind was still blowing from the WD surface, but the wind near the binary system was obscured in the photosphere until day $\sim$6.\\

    \item
    Between roughly days 6 and 8, the narrow components of H$\alpha$ and He \textsc{ii} 4686, 
    mainly originating from the nova wind near the binary system, became clearly visible.
    Soft X-ray photons from the steady H-burning on the WD surface were also faintly visible on day $\sim$8.
    These facts suggest that the ejecta photosphere receded close to the binary system.\\

    \item
    Subsequently, the ejecta photosphere receded inside the binary system.
    The primary optical eclipse was first observed on day $\sim$11.
    This first eclipse might be due to the receding ejecta photosphere eclipsed by the secondary,
    with the optical brightness of the photosphere having decreased sufficiently for the optical eclipse to be detectable.
    It should be noted that the accretion disk was still expanded due to the nova wind and might be outside the photosphere.
    However, the mass-loss rate driven by the nova wind was still high,
    causing a large part of the energy flux from the WD to be self-absorbed by the wind itself.
    Therefore, the expanded accretion disk was not sufficiently irradiated by the hotter WD,
    and it was not luminous enough to show the eclipses in the optical range.\\

    \item     
    The mass-loss rate driven by the nova wind gradually became lower.
    During the optical plateau stage, roughly between days 14 and 24, 
    the WD surface, sustaining the steady H-burning, fully emerged from the ejecta photosphere, and the soft X-ray photons became evident.
    The expanded accretion disk was fully irradiated by the emerging hotter WD surface, resulting in a main light source in the optical range.
    As a result, the disk began to show the optical eclipses, with the eclipse width larger than that in quiescence. 
    The outgoing disk surface flow, as well as the nova wind near the binary system, might contribute to the narrow components. \\

    \item 
    Just after the optical plateau stage ended on day $\sim$24,
    the steady H-burning on the WD surface gradually weakened, and the nova wind became too weak to expand the accretion disk.
    Eventually, the accretion disk drastically shrank to the tidal truncation radius
    within just a few orbital periods (i.e., a few days) after the end of the optical plateau stage.
    Then, the wind became even weaker or stopped, and the narrow components originating from the nova wind disappeared on day $\sim$40,
    with the disk line profiles persisting.
    Finally, the steady H-burning on the WD surface completely finished, and the detected soft X-ray photons became faint on day $\sim$50.\\
    
    \end{enumerate}

\subsection{Disk radius in quiescence}
    
    Based on the DP of the H$\alpha$ line profile in the 2024 quiescent spectrum of U Sco, 
    we can roughly estimate the outer radius of the accretion disk in quiescence.
    Generally, the observed splitting of the double-peaked emission line corresponds to a velocity 
    which is 70\% $\pm$ 10\% of the Keplerian velocity at the outer edge of the accretion disk \citep{wade1988}.
    Thus, we can simply establish the following equation [see also equation (2) of \citealp{casares2016}]:

    \begin{eqnarray}
        \frac{\mathrm{DP}}{2} = \beta\left(\frac{GM_1}{R_\mathrm{disk}}\right)^{1/2}\sin i .
    \end{eqnarray}    

    \noindent
    The parameter $\beta$ represents the fraction by which the outer disk material deviates from its Keplerian velocity,
    and here, we set $\beta = 0.7$.\footnote{
    \ Based on \citet{Rtid}, a test particle on a periodic orbit at the tidal truncation radius in U Sco has a velocity of $\sim$70\% 
    of the expected Keplerian velocity of a circular orbit at the corresponding radius in the assumed gravity of only the primary WD.}
    $G$ represents the gravitational constant.
    The primary WD mass $M_1$ is estimated to be $1.37 \pm 0.01\ \MO$ \citep{hachisu2000}.
    The orbital inclination $i$ is calculated to be $82.7\degree \pm 2.9\degree$ \citep{thoroughgood2001}.
    Substituting the DP values obtained in subsection \ref{subsec:quie}, 
    the outer radius of the accretion disk is calculated to be $R_\mathrm{disk} = 2.32(5)\ R_1^{*}$,
    where the primary's Roche volume radius $R_1^{*} = 2.64\ \RO$ for U Sco \citep{muraoka2024}.

    This result is implausible and likely an overestimate, considering that
    this value is comparable to the orbital separation $a = 6.33\ \RO = 2.40\ R_1^{*}$
    and is much larger than the tidal truncation radius $R_\mathrm{tid} = 0.85\ R_1^{*}$ \citep{muraoka2024}.
    Meanwhile, \citet{hachisu2000b} have modeled the quiescent optical light curve of U Sco and concluded that 
    the outer disk radius is estimated to be $R_\mathrm{disk} = 0.7\ R_1^{*}$ in quiescence.
    Usually, the disk radius estimated from the double-peaked line profile is larger than
    that estimated from other techniques (\citealp{wade1988}; \citealp{marsh1988}; subsection 2.7.4 of \citealp{warner}).
    Apart from the sub-Keplerian motion, \citet{marsh1988} has pointed out that 
    there is no definite outer cut-off in the distribution of the line emission.

    In the case of a thick accretion disk viewed at a high inclination, 
    lines-of-sight traverse a wide range of radii in passing through the accretion disk,
    and Keplerian shear broadening has a dominant influence on the local radiative transfer (\citealp{horne1986}; \citealp{horne1995}).
    They have stated that this shear results in the V-shaped central trough in the double-peaked line profile
    \citep[e.g., H$\alpha$ and H$\beta$ line profiles of WZ Sge in][]{skidmore2000}.
    \citet{rutkowski2016} have studied the optical spectra of high-inclination dwarf nova V2051 Oph, suggesting that 
    circumstances dominated by Keplerian shear broadening can significantly distort the inferred measurements 
    of the outer disk radius based on the double-peak separation.

    A similar scenario can be considered for U Sco.
    As mentioned before, the orbital inclination is calculated to be $i = 82.7\degree \pm 2.9\degree$, and
    the scale height at the outer edge of the accretion disk is estimated to be $H/R \sim 0.3$ in quiescence \citep{hachisu2000b}.
    This condition satisfies $\tan i > R / H$, suggesting that Keplerian shear broadening is likely to play a significant role \citep{horne1986}.
    Therefore, it is considered to be quite challenging 
    to derive the accurate outer radius of the accretion disk in quiescence from the double-peaked line profile.

\section{Summary}\label{sec05:sum} 

    We present the spectral evolution of the H$\alpha$ and He \textsc{ii} 4686 line profiles
    during the U Sco 2022 eruption over 22 d following the optical peak,
    and the quiescent H$\alpha$ line profile on day $\sim$794 after the U Sco 2022 eruption.
    The H$\alpha$ line profile shows a narrow single peak on day 16.41,
    which is different from a triple-peaked profile observed on day 17 after the U Sco 1999 eruption.
    We attribute this difference to the secondary located between the observer and the primary WD at the onset of the 1999 eruption,
    potentially altering the mass distribution of the ejecta or shocked shell on the observer's side.
    Moreover, we demonstrate that the single-peaked narrow components of H$\alpha$ and He \textsc{ii} 4686 
    appeared almost simultaneously between roughly days 6 and 8, 
    preceding the onset of the optical eclipses of the accretion disk observed after day 11 during the U Sco 2022 eruption.
    We propose that the nova wind near the binary system may be the primary origin of these narrow components 
    and even remained active several days after the nova eruption with a velocity of approximately 1000 $\mathrm{km\ s}^{-1}$,
    likely driving the expansion of the accretion disk until the end of the optical plateau stage. 
    While the contribution of the rotating accretion disk might be dominated by that of the nova wind in the H$\alpha$ line profile,
    we might also observe the outward surface flow from the expanded disk contributing to these narrow features during the optical plateau stage.
    The double-peak separation of the H$\alpha$ line profile in quiescence could not provide a reliable estimate for the outer disk radius,
    possibly due to the thickness of the accretion disk observed at a high inclination.
    
\section*{Supplementary data} 

    The following supplementary data is available at PASJ online.

    E-figures 1--4 and e-table 1.

\begin{ack}

    This work was supported by many BAA, AAVSO and ARAS observers.
    We acknowledge with thanks the variable star observations from the AAVSO International Database contributed by observers worldwide and used in this research.
    All ARAS observers who contribute to this program are warmly commended for their tireless efforts.
    This work is partially supported by the Optical and Infrared Synergetic Telescopes for Education and Research (OISTER) program funded by the MEXT of Japan.
    This work also makes use of the photometric data from \citet{muraoka2024}.
    We would like to express our gratitude to the observers for their contributions.
    We are grateful to T. Iijima for providing us with valuable comments regarding the spectral differences between the U Sco 1999 and 2022 eruptions.
    We also thank F. Iwamuro for developing a data reduction pipeline for KOOLS-IFU.
    This work is partly supported by JSPS KAKENHI Grant Number 20K14521.
    This work is supported by Slovak Research and Development Agency under contract No. APVV-20-0148.
    
\end{ack}

\section*{Data availability} 

    The photometric data and codes for the eclipse analysis provided by \citet{muraoka2024} are available at <https://doi.org/10.14989/292192>.
    Our spectroscopic data can be found at <https://doi.org/10.14989/292193>,
    both in the Kyoto University Research Information repository (KURENAI).\footnote{
    \ <https://repository.kulib.kyoto-u.ac.jp/dspace/?locale=en>.}






\bibliographystyle{pasjlike}
\bibliography{cvs}

\newcommand{\noop}[1]{}
\begin{thebibliography}{}

\bibitem[Abril-Pla  et~al.(2023)]{PyMC5}
  Abril-Pla, O., {et~al.}\ 2023, PeerJ Comput. Sci. 9:e1516 DOI 10.7717/peerj-cs.1516

\bibitem[Anupama  et~al.(2013)]{anupama2013}
  Anupama, G.~C., {et~al.}\ 2013, A\&A, 559, A121

\bibitem[Azzollini et~al.(2023)]{azzollini2023}
  Azzollini, A., Shore, S.~N., Kuin, P., \& Page, K.~L.\ 2023, A\&A, 674, A139

\bibitem[Brandi et~al.(2009)]{brandi2009}
  Brandi, E., Quiroga, C., Mikołajewska, J., Ferrer, O.~E., \& García, L.~G.\ 2009, A\&A, 497, 815

\bibitem[Casares(2015)]{casares2015}
  Casares, J.\ 2015, ApJ, 808, 80

\bibitem[Casares(2016)]{casares2016}
  Casares, J.\ 2016, ApJ, 822, 99

\bibitem[Chomiuk et~al.(2021)]{chomiuk2021}
  Chomiuk, L., Metzger, B.~D., \& Shen, K.~J.\ 2021, Annual Review of Astronomy and Astrophysics, 59, 391

\bibitem[Darnley  et~al.(2017)]{darnley2017}
  Darnley, M.~J., {et~al.}\ 2017, ApJ, 849, 96

\bibitem[Drake and Orlando(2010)]{drake2010}
  Drake, J.~J., \& Orlando, S.\ 2010, ApJLett, 720, L195

\bibitem[Eggleton(1983)]{eggleton1983}
  Eggleton, P.~P.\ 1983, ApJ, 268, 368

\bibitem[Evans et~al.(2023)]{evans2023}
  Evans, A., Banerjee, D. P.~K., Woodward, C.~E., Geballe, T.~R., Gehrz, R.~D., Page, K.~L., \& Starrfield, S.\ 2023, MNRAS, 522, 4841

\bibitem[Figueira et~al.(2025)]{figueira2025}
  Figueira, J., José, J., Cabezón, R., \& García-Senz, D.\ 2025, A\&A, 693, A209

\bibitem[Figueira et~al.(2018)]{figueira2018}
  Figueira, J., José, J., García-Berro, E., Campbell, S.~W., García-Senz, D., \& Mohamed, S\ 2018, A\&A, 613, A8

\bibitem[Frank et~al.(2002)]{frank2002}
  Frank, J., King, A., \& Raine, D.\ 2002, Accretion Power in Astrophsics (Cambridge: Cambridge University Press)

\bibitem[Gallagher and Starrfield(1978)]{Nova}
  Gallagher, J.~S., \& Starrfield, S.\ 1978, Annual Review of Astronomy and Astrophysics, 16, 171

\bibitem[Gelman et~al.(2013)]{PSR}
  Gelman, A., Carlin, J.~B., Stern, H.~S., Dunson, D.~B., Vehtari, A., \& Rubin, D.~B.\ 2013, Baysian Data Analysis Third Edition (Chapman \& Hall/CRC Press)

\bibitem[Hachisu and Kato(2003)]{hachisu2003}
  Hachisu, I., \& Kato, M.\ 2003, ApJ, 588, 1003

\bibitem[Hachisu and Kato(2006)]{hachisu2006_2}
  Hachisu, I., \& Kato, M.\ 2006, ApJS, 167, 59

\bibitem[Hachisu and Kato(2022)]{hachisu2022}
  Hachisu, I., \& Kato, M.\ 2022, ApJ, 939, 1

\bibitem[Hachisu and Kato(2023)]{hachisu2023}
  Hachisu, I., \& Kato, M.\ 2023, ApJ, 953, 78

\bibitem[Hachisu et~al.(2000a)]{hachisu2000}
  Hachisu, I., Kato, M., Kato, T., \& Matsumoto, K.\ 2000a, ApJ, 528, L97

\bibitem[Hachisu et~al.(2000b)]{hachisu2000b}
  Hachisu, I., Kato, M., Kato, T., Matsumoto, K., \& Nomoto, K.\ 2000b, ApJ, 534, L189

\bibitem[Hachisu et~al.(2024)]{hachisu2024}
  Hachisu, I., Kato, M., \& Matsumoto, K.\ 2024, ApJ, 965, 49

\bibitem[Hachisu et~al.(2025)]{hachisu2025}
  Hachisu, I., Kato, M., \& Walter, F.~M.\ 2025, ApJ, 980, 142

\bibitem[Hart  et~al.(2023)]{ASN_3}
  Hart, K., {et~al.}\ 2023, arXiv:2304.03791

\bibitem[Hellier(2001)]{hellier}
  Hellier, C.\ 2001, Cataclysmic Variable Stars (Berlin: Springer)

\bibitem[Henze  et~al.(2018)]{henze2018}
  Henze, M., {et~al.}\ 2018, ApJ, 857, 68

\bibitem[Hoffman and Gelman(2014)]{NUTS}
  Hoffman, M.~D., \& Gelman, A.\ 2014, Journal of Machine Learning Research, 15, 1593

\bibitem[Horne(1995)]{horne1995}
  Horne, K.\ 1995, A\&A, 297, 273

\bibitem[Horne and Marsh(1986)]{horne1986}
  Horne, K., \& Marsh, T.~R.\ 1986, MNRAS, 218, 761

\bibitem[Iijima(2002)]{iijima2002}
  Iijima, T.\ 2002, A\&A, 387, 1013

\bibitem[Kato and Hachisu(1994)]{mkato1994}
  Kato, M., \& Hachisu, I.\ 1994, ApJ, 437, 802

\bibitem[Kato et~al.(2022)]{mkato2022}
  Kato, M., Saio, H., \& Hachisu, I.\ 2022, ApJLett, 935, L15

\bibitem[Kato(2023)]{tkato2023}
  Kato, T.\ 2023, Variable Star Bulletin, No. 115

\bibitem[Kurita  et~al.(2020)]{seimei}
  Kurita, M., {et~al.}\ 2020, PASJ, 72, 48

\bibitem[Leahy and Leahy(2015)]{leahy2015}
  Leahy, D.~A., \& Leahy, J.~C.\ 2015, Computational Astrophysics and Cosmology, 2, 4

\bibitem[Leibowitz et~al.(1992)]{leibowitz1992}
  Leibowitz, E.~M., Mendelson, H., Mashal, E., Prialnik, D., \& Seitter, W.~C.\ 1992, ApJ, 385, L49

\bibitem[McClintock et~al.(1975)]{BOWEN}
  McClintock, J.~E., Canizares, C.~R., \& Tarter, C.~B.\ 1975, ApJ, 198, 641

\bibitem[Marsh(1988)]{marsh1988}
  Marsh, T.~R.\ 1988, MNRAS, 231, 1117

\bibitem[Mason et~al.(2012)]{mason2012}
  Mason, E., Ederoclite, A., Williams, R.~E., Della~Valle, M., \& Setiawan, J.\ 2012, A\&A, 544, A149

\bibitem[Mason and Munari(2014)]{mason2014}
  Mason, E., \& Munari, U.\ 2014, A\&A, 569, A84

\bibitem[Matsubayashi  et~al.(2019)]{KIF}
  Matsubayashi, K., {et~al.}\ 2019, PASJ, 71, 102

\bibitem[Maxwell et~al.(2014)]{maxwell2014}
  Maxwell, M.~P., Rushton, M.~T., \& Eyres, S. P.~S.\ 2014, ASP\ Conf.\ Ser.\, 490, 205

\bibitem[Munari et~al.(2010)]{munari2010}
  Munari, U., Dallaporta, S., \& Castellani, F.\ 2010, IBVS, 5930, 1

\bibitem[Muraoka  et~al.(2024)]{muraoka2024}
  Muraoka, K., {et~al.}\ 2024, PASJ, 76, 293

\bibitem[Paczyński(1977)]{Rtid}
  Paczyński, B.\ 1977, ApJ, 216, 822

\bibitem[Page  et~al.(2022)]{page2022_3}
  Page, K.~L., {et~al.}\ 2022, MNRAS, 514, 1557

\bibitem[Retter et~al.(1997)]{retter1997}
  Retter, A., Leibowitz, E.~M., \& Ofek, E.~O.\ 1997, MNRAS, 286, 745

\bibitem[Rudy et~al.(2023)]{rudy2023}
  Rudy, R.~J., Subasavage, J.~P., Mauerhan, J.~C., \& Nofi, L.\ 2023, RNAAS, 7, 26

\bibitem[Rutkowski et~al.(2016)]{rutkowski2016}
  Rutkowski, A., Waniak, W., Preston, G., \& Pych, W.\ 2016, MNRAS, 463, 3290

\bibitem[Sala and Hernanz(2005)]{sala2005}
  Sala, G., \& Hernanz, M.\ 2005, A\&A, 439, 1061

\bibitem[Schaefer(2022)]{schaefer2022}
  Schaefer, B.~E.\ 2022, MNRAS, 516, 4497

\bibitem[Schaefer  et~al.(2011)]{schaefer2011}
  Schaefer, B.~E., {et~al.}\ 2011, ApJ, 742, 113

\bibitem[Schmidtobreick et~al.(2003)]{schmidtobreick2003}
  Schmidtobreick, L., Tappert, C., Bianchini, A., \& Mennickent, R.~E.\ 2003, A\&A, 410, 943

\bibitem[Shappee  et~al.(2014)]{ASN_1}
  Shappee, B.~J., {et~al.}\ 2014, ApJ, 788, 48

\bibitem[Skidmore et~al.(2000)]{skidmore2000}
  Skidmore, W., Mason, E., Howell, S.~B., Ciardi, D.~R., Littlefair, S., \& Dhillon, V.~S.\ 2000, MNRAS, 318, 429

\bibitem[Sokoloski et~al.(2008)]{sokolovsky2008}
  Sokoloski, J.~L., Rupen, M.~P., \& Mioduszewski, A.~J.\ 2008, ApJ, 685, L137

\bibitem[Starrfield et~al.(1972)]{TNR}
  Starrfield, S., Truran, J.~W., Sparks, W.~M., \& Kutter, G.~S.\ 1972, ApJ, 176, 169

\bibitem[Teyssier(2019)]{ARAS}
  Teyssier, F.\ 2019, Contr.\ of\ the\ Astron.\ Obs.\ Skalnat\'e Pleso, 49, 217

\bibitem[Thoroughgood et~al.(2001)]{thoroughgood2001}
  Thoroughgood, T.~D., Dhillon, V.~S., Littlefair, S.~P., Marsh, T.~R., \& Smith, D.~A.\ 2001, MNRAS, 327, 1323

\bibitem[Wade and Horne(1988)]{wade1988}
  Wade, R.~A., \& Horne, K.\ 1988, ApJ, 324, 411

\bibitem[Warner(1995)]{warner}
  Warner, B.\ 1995, Cataclysmic Variable Stars (Cambridge: Cambridge University Press)

\bibitem[Wolf et~al.(2013)]{wolf2013}
  Wolf, W.~M., Bildsten, L., Brooks, J., \& Paxton, B.\ 2013, ApJ, 777, 136

\bibitem[Worters et~al.(2007)]{worters2007}
  Worters, H.~L., Eyres, S. P.~S., Bromage, G.~E., \& Osborne, J.~P.\ 2007, MNRAS, 379, 1557

\bibitem[Worters et~al.(2010)]{worters2010}
  Worters, H.~L., Eyres, S. P.~S., Rushton, M.~T., \& Schaefer, B.\ 2010, IAU\ Circ., 9114, 1

\bibitem[Yamanaka  et~al.(2010)]{yamanaka2010}
  Yamanaka, M., {et~al.}\ 2010, PASJ, 62, L37

\bibitem[Zamanov et~al.(2006)]{zamanov2006}
  Zamanov, R., Boër, M., Le~Coroller, H., \& Panov, K.\ 2006, IBVS, 5733, 1

\end{thebibliography}

\end{document}